**Closure in turbulence from first principles**.

Michail Zak
Senior Research Scientist (Emeritus)
Jet Propulsion Laboratory California Institute of Technology
Pasadena, CA 91109, USA.

It has been recently demonstrated, [3], that according to the principle of release constraints, absence of shear stresses in the Euler equations must be compensated by additional degrees of freedom, and that led to a Reynolds-type enlarged Euler equations (EE equations) with a doublevalued velocity field that ***do not require any closures***. In the first part of the paper, the theory is applied to turbulent mixing and illustrated by propagation of mixing zone triggered by a tangential jump of velocity. A comparison of the proposed solution with the Prandtl's solution is performed and discussed. In the second part of the paper, a semi-viscous version of the Navier-Stokes equations is introduced. The ***model does not require any closures*** since the number of equations is equal to the number of unknowns. Special attention is paid to transition from laminar to turbulent state. The analytical solution for this transition demonstrates the turbulent mean velocity profile that qualitatively similar to celebrated logarithmic law.

**1. Introduction.**

During several centuries, inviscid incompressible fluid – the simplest model of a continuum - enjoyed an unprecedented success being considered as the most elegant branch of continuum mechanics. It stimulated progress in theory of conformal transformations, and theory of harmonic functions thereby transferring new mathematical techniques to theory of elasticity as well as to other types of continua. However more detailed studies of the solutions of Euler's equations demonstrated fundamental inconsistencies of the model. The first inconsistency was associated with zero-drag-paradox proved by D'Alembert in 1752, [1]. Recent numerical study performed by Yudovich, [2], reports a general result that solutions of Euler's equations *inherently unstable* and in a finite time they become stochastic. Such flaws of the Euler's model triggered "escape" to Navier-Stokes equations of viscous fluid. However the viscose model has problems of its own: existence and uniqueness of solutions of the Navier-Stokes equations have not yet been proven, the solutions loss stability at regions of supercritical Raynolds numbers, etc. On the top of that, neither Euler, nor Navier-Stokes equations are capable to describe turbulent motions. It is interesting to notice that although turbulence is usually associated with the Navier-Stokes rather than the Euler's equations, actually in developed turbulence viscosity plays vanishingly small role, and that was the main motivation to revisit the Euler's model.

   The source of the problems with both Euler and Navier-Stokes equations was found and discussed in our recent publication, [3]. It has been demonstrated that according to the principle of release of constraints, absence of shear stresses must be compensated by additional degrees of freedom, and that leads to enlarged Euler's equations (EE equations) with a doublevalued velocity field. Analysis of coupled mean/fluctuation EE equations showed that fluctuations stabilize the whole system generating elastic shear waves. This opens the way to apply EE equations for postinstability regions of the Navier –Stokes equations instead of the Reynolds equations, [4], and thereby to avoid a closure problem.

In this paper, thorough derivation of the EE equations from the principle of virtual work is elaborated, and invariants such as first integrals, as well as characteristics speeds are formulated. Special attention is paid to a semi-viscous version of the Navier-Stokes equation that is also based upon doublevalued velocity field.

## 2. Instability in dynamics.

Any mathematical model of a continuum should be tested for three properties: existence, uniqueness and stability of its solutions. However, none of these properties are physical invariants since they depend upon the class of functions in which the solution is sought. As an example, consider a vertical, ideally flexible filament with a free lower end suspended in the gravity field. As shown in [5], the unique stable solution exists in the class of functions satisfying the Lipchitz condition. However despite its "nice" mathematical properties, this solution is in contradiction with experiments: the cumulative effect – snap of a whip – is lost. At the same time, the removal of the Lipchitz conditions leads to non-unique unstable solutions that perfectly describe the snap of a whip. Another example, [6], illustrates the dependence of stability of the solution upon the frame of reference: consider an inviscid stationary flow with a *smooth* velocity field

$$v_x = A\sin z + C\cos y, \ v_y = B\sin x + A\cos z, \ v_z = C\sin y + b\cos x$$

Surprisingly, the trajectories of this flow are unstable (Lagrangian turbulence). It means that this flow is stable in the Eulerian coordinates, but is unstable in the Lagrangian coordinates.

It is important to distinguish short- and long-term instabilities. Short-term instability occurs when the system has alternative stable states (an inverted pendulum); it is characterized by bounded deviations of position coordinates whose change affects the energy of the system, and therefore this type of instability does not require a model modification. The long-term instability occurs when the system does not have an alternative stable state. Such instability can involve only ignorable coordinates since these coordinates do not affect the energy of the system. That is why the long-term instability, from physical viewpoint, can be associated with chaos, and from mathematical viewpoint – with the loss of smoothness, or with the loss of differentiability. And that is why the long-term instability requires a modification of the model. Since the Euler's model of inviscid fluid abounds with chaotic instabilities with no alternative stable states, modification of this model is the main subject of this paper. It should be recalled that the first step in this direction was made in our recent publication [3].

It should be mentioned that the long-term instability is subdivided, at least, in two different groups: Lyapunov instability that is associated with unbounded growth of some selected modes, and Hadamard instability that results from degeneration of a hyperbolic PDE into an elliptic PDE, [5], while all the modes grow unboundedly. That is why the Hadamard instability is based upon local relationships that do not explicitly depend upon boundary conditions. In addition to that, in case of

Hadamard instability, an infinitesimal initial disturbance becomes finite in finite time period, [5], while in Lyapunov instability case this period must be infinite.

In the Euler's model of inviscid fluid, both type of long-term instability occurs: vortices are Lyapunov -unstable, [2], and tangential discontinuities of velocity are Hadamard-unstable, [3]. As demonstrated in [3], both of these instabilities are suppressed by fluctuations in EE model.

### 3. Postinstability model of inviscid fluid.

The concept of multivaluedness of the velocity field in inviscid fluid was introduced in [3] as a way of removing inherent instability by enlarging the class of functions in which the solutions are sought. As shown there, exceptionally in inviscid fluid, multivaluedness of the velocity field does not cause unbounded stresses since the viscosity is zero. In this section we will elaborate a formalism of derivation of the modified model using the principle of virtual work.

Let us define a postinstability state of a fluid as following

$$\mathbf{v}(\mathbf{r}_2) \neq \mathbf{v}(\mathbf{r}_1), \text{ if } \mathbf{r}_2 = \mathbf{r}_1 \tag{1}$$

where **v** is the velocity, and **r** is the position vector of a point in space.

This means that two different particles with different velocities can appear at the same point of space without causing unbounded stresses. Actually this is an idealization of the condition

$$\mathbf{v}(\mathbf{r}_2) \neq \mathbf{v}(\mathbf{r}_1) \text{ if } \mathbf{r}_2 \to \mathbf{r}_1 \tag{2}$$

that takes place in turbulent motion. It is not a coincidence that the same condition (2) is observed in the course of Hadamard's instability of tangential jump of velocity that was analyzed and discussed in [3]. In addition to that, the multivaluedness (1) can be imposed by boundary conditions with sharp angles or cones as a result of inviscid fluid slip at the boundary.

Let us express the condition (1) in terms of variations of virtual velocities $\hat{\mathbf{v}}$

$$\delta\hat{\mathbf{v}} \neq 0 \quad at \quad \delta\mathbf{r} = 0 \tag{3}$$

and compare it with the condition

$$\delta\hat{\mathbf{v}} = 0 \quad at \quad \delta\mathbf{r} = 0 \tag{4}$$

The last condition states that if the position of a point of the fluid is fixed, the fluid velocity at this point is also fixed, and that defines a singlevalued function $\mathbf{v}(\mathbf{r})$. In this context, the condition (3) states that at a fixed point *r* the fluid velocity can have many different values $\mathbf{v}_1, \mathbf{v}_2, ... \mathbf{v}_n$, and that defines a *n*-valued function

$$\mathbf{v}(\mathbf{r}, \xi), \quad \xi = 1, 2, ... n, \quad \mathbf{v}_i = \mathbf{v}(\mathbf{r}, \xi_i) \tag{5}$$

In general, the maltivaluedness parameters $\xi$ are non-necessarily discrete numbers: they can form a continuum as well in a close interval, for instance

$$0 \leq \xi \leq 1 \tag{6}$$

For derivation of the governing equation of a continuum, we will start with the principle of virtual work

$$\int_V (\rho \mathbf{a} - \mathbf{F}) \cdot \delta \hat{\mathbf{v}} \, dV = 0 \tag{7}$$

where V is an arbitrary volume in space occupied by the medium, $\hat{\mathbf{v}}$ is a virtual velocity, $\mathbf{a}$ is the acceleration, and $\rho$ is the density.

For a continuum with a singlevalued velocity field, the condition (4) is true. We will express it in terms of the velocity tensor-gradient

$$\delta \nabla \hat{\mathbf{v}} = 0, \tag{8}$$

Let us now multiply the equality (8) by a Lagrange multiplier represented by an arbitrary tensor $(T)^T$ of the same rank and of dimensionality of stress

$$(T)^T \cdot\cdot \, \delta \nabla \hat{\mathbf{v}} = 0 \tag{9}$$

where $(T)^T$ denotes the transpose of T.

Taking into account the identity

$$\nabla \cdot [(A)^T \cdot \mathbf{r}] = \mathbf{r} \cdot \nabla \cdot (A)^T + (A)^T \cdot\cdot (\nabla \mathbf{r})^T \tag{10}$$

where A and $\mathbf{r}$ are arbitrary tensor and vector, and the Gauss theorem

$$\int_V \nabla \cdot \mathbf{r} \, dV = \oint_\Sigma \mathbf{r} \, dS \tag{11}$$

then integrating the constraint (9) over the volume V, and adding term by term to the equality (7) one obtains

$$\int_V (\rho \mathbf{a} - \mathbf{F} - \nabla \cdot T) \cdot \delta \hat{\mathbf{v}} \, dV + \oint_\Sigma (\mathbf{F} - T \cdot \mathbf{n}) \cdot \delta \hat{\mathbf{v}} \, d\sigma = 0 \tag{12}$$

whence because of independence of variations $\delta \hat{\mathbf{v}}$

$$\rho \mathbf{a} = \nabla \cdot (devT - pE) + \mathbf{F} \tag{13}$$

$$(devT - pE) \cdot \mathbf{n} = \mathbf{F}_S$$

Here $\Sigma$ is the surface bounding the volume V, $\mathbf{n}$ is the unit normal to this surface, and $\mathbf{F}, \mathbf{F}_S$ are the volume and surface external forces, and E is the unit tensor.

Eqs. (13) present the governing equations of a classical continuum with the corresponding boundary conditions. In our setting, the stress tensor T plays the role of a reaction to the kinematical constraint (8). In other words, the requirement of a singlevaluedness of the velocity field is equivalent to existence of the stress tensor.

Let us now derive the governing equations for an incompressible inviscid fluid as a particular case of Eqs. (13). It turns out that this is not as trivial as it seems. First decompose the velocity tensor-gradient as well as the stress tensor into spherical and deviatoric parts

$$\nabla \hat{\mathbf{v}} = \frac{1}{3} (\nabla \cdot \hat{\mathbf{v}}) E + dev(\nabla \hat{\mathbf{v}}) \tag{14}$$

$$T = -\frac{1}{3}pE + dev(\nabla \hat{v}) \tag{15}$$

where *p* is the spherical part of the tensor T (pressure).
Then the equality (9) can be also decomposed

$$p\delta(\nabla \cdot \hat{v}) = 0 \tag{16}$$

$$dev(T)^T \cdot\cdot \delta(dev\nabla\hat{v}) = 0 \tag{17}$$

and after similar transformations, one arrived at the governing equations equivalent to Eqs. (13)

$$\rho \mathbf{a} = \nabla \cdot (-p) + \mathbf{F} \tag{18}$$

$$-p\mathbf{n} = \mathbf{F}_\Sigma \tag{19}$$

Since in inviscid fluid, by definition, the tangential stresses are zero

$$devT = 0 \tag{20}$$

one can substitute Eq. (20) into Eqs. (18) and (19) and arrive at the Euler's equations

$$\rho \mathbf{a} = \nabla \cdot (-p) + \mathbf{F} \tag{21}$$

$$-p\mathbf{n} = \mathbf{F}_\Sigma \tag{22}$$

However, these equations are **incomplete**! Indeed, if one turns to the equality (17), it becomes clear that Eq. (20) open up a possibility of multivaluedness of the deviatoric components of the velocity field

$$\delta(dev\nabla\hat{v}) \neq 0 \tag{23}$$

At the same time, the divergency of velocity remains singlevalued since

$$\delta(\nabla \cdot \hat{v}) = 0 \quad if \ p \neq 0 \tag{24}$$

It should be noticed that the equality (17) actually represents the principle of release of constraints: each constraint suppresses a certain degree of freedom, and if this constraint is removed, the corresponding degree of freedom should be released. In our case the constraint is the existence of non-zero tangential stresses, and the suppressed degree of freedom is the maltivaluedness of the deviatoric components of the velocity field; hence as soon as the tangential stresses vanish, the multivaluedness reappears.
The same result has been obtained in our previous publication [3] based upon requirement that stresses in a continuum must be bounded.
Actually incompleteness of the Euler's equations explains their inherent instability.
**4. Enlarged Euler's equations, (EE equations).**
The result formulated above was interpreted in [3] as following: an inviscid fluid can be considered as a result of superposition of *n* physically identical, but kinematically different continua, Fig.1. In case of incompressible fluid there are *n+1* governing equations with respect to *n+1* independent variables

$$\frac{\partial \mathbf{v}_i}{\partial t} + \mathbf{v}_i \nabla \mathbf{v}_i = \frac{1}{\rho}(-\nabla p + \mathbf{F}), \quad i = 1,2,...n \tag{25}$$

coupled via the mass conservation equation

$$\nabla \cdot \sum_{i=1}^{n} \mathbf{v}_i = 0 \tag{26}$$

Introducing the velocity $\bar{\mathbf{v}}$ of the "center of inertia" of the $n$ particles superimposed at the same point of space (an analog of the classical velocity), and the fluctuations with respect to the "center of inertia" $\tilde{\mathbf{v}}_i, i = 1, 2, \ldots n$, one obtains the following decomposition

$$\mathbf{v}_i = \bar{\mathbf{v}} + \tilde{\mathbf{v}}_i \qquad \bar{v} = \frac{1}{n} \sum_{i=1}^{n} v_i \tag{27}$$

Obviously

$$\sum_{i=1}^{n} \tilde{v}_i = 0 \tag{28}$$

Exploring these decompositions, Eqs. (25) and (26) can be presented in the form

$$\frac{\partial \bar{\mathbf{v}}}{\partial t} + \bar{\mathbf{v}} \nabla \bar{\mathbf{v}} + \sum_{i=1}^{n} \tilde{\mathbf{v}}_i \nabla \tilde{\mathbf{v}}_i = \frac{1}{\rho}(-\nabla p + \mathbf{F}) \tag{29}$$

$$\frac{\partial \tilde{\mathbf{v}}_i}{\partial t} + \bar{\mathbf{v}} \nabla \tilde{\mathbf{v}}_i + \tilde{\mathbf{v}}_i \nabla \bar{\mathbf{v}} - \sum_{i=1}^{n} \tilde{\mathbf{v}}_i \nabla \tilde{\mathbf{v}}_i = 0, \quad i = 1, 2, \ldots n \tag{30}$$

$$\nabla \cdot \sum_{i=1}^{n} \bar{\mathbf{v}} = 0 \tag{31}$$

Eqs.(29),(30),and (31) form a **closed** system of $n+1$ vector and one scalar equations with respect to $n+1$ vector and one scalar unknowns.
In this paper as in [3], we will concentrate on a doublevalued model.
The doublevalued version of this system, with the notations

$$\mathbf{v} = \bar{\mathbf{v}} \pm \tilde{\mathbf{v}}, \quad \bar{\mathbf{v}} = \frac{1}{2}(\mathbf{v}_1 + \mathbf{v}_2), \quad \tilde{\mathbf{v}} = \pm \frac{1}{2}(\mathbf{v}_1 - \mathbf{v}_2) \tag{32}$$

is simplified to the following

$$\frac{\partial \bar{\mathbf{v}}}{\partial t} + \bar{\mathbf{v}} \nabla \bar{\mathbf{v}} + \tilde{\mathbf{v}} \nabla \tilde{\mathbf{v}} = \frac{1}{\rho}(-\nabla p + \mathbf{F}) \tag{33}$$

$$\frac{\partial \tilde{\mathbf{v}}}{\partial t} + \bar{\mathbf{v}} \nabla \tilde{\mathbf{v}} + \tilde{\mathbf{v}} \nabla \bar{\mathbf{v}} = 0 \tag{34}$$

$$\nabla \cdot \bar{\mathbf{v}} = 0 \tag{35}$$

It slightly resembles the Reynolds equations, but there are several fundamental differences emphasized in [3]: Firstly the EE systems are closed, i.e. the number of unknowns is equal to the number of equations, and that eliminates the closure problem. Secondly the EE system do not include the continuity equation for fluctuations. These differences follow from the fact that the Reynolds velocity field, strictly speaking, is single-valued since the stress tensor of the Navier-Stokes equations does not have zero components. In other words, the condition (1) for the Reynolds velocity field should be replaced by a weaker condition (2)

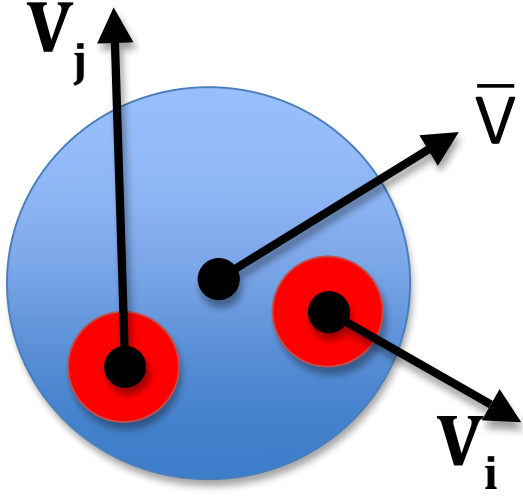

**Figure 1. Superposition of two different particles at the same point of space.**

**5. Integral form of the governing equations.**
In this paper, as in [3], we will deal only with the doublevalued model since all the specific features of the EE equations become more transparent. In order to study propagation of tangential jumps of velocities in inviscid model of turbulence, we will formulate the governing equations (33)-(35) in the integral form. For that purpose we will apply the laws of conservation of momentum and energy first to each half-particle separately

$$(\int_V \rho \mathbf{v_i} dV)_{t=t_2} - (\int_V \rho \mathbf{v_i} dV)_{t=t_1} = -\int_{t_1}^{t_2} (\int_S p\mathbf{n} dS) dt, \quad i=1,2. \tag{36}$$

$$(\int_V \rho \frac{\mathbf{v_i} \cdot \mathbf{v_i}}{2} dV)_{t=t_2} - (\int_V \rho \frac{\mathbf{v_i} \cdot \mathbf{v_i}}{2} dV)_{t=t_1} = -\int_{t_1}^{t_2} (\int_S p\mathbf{v_i} \cdot \mathbf{n} dS) dt, \quad i=1,2. \tag{37}$$

After returning to mean and fluctuation velocities, these equations take the form, respectively

$$(\int_V \rho \bar{\mathbf{v}} \, dV)_{t=t_2} - (\int_V \rho \bar{\mathbf{v}} dV)_{t=t_1} = -\int_{t_1}^{t_2} (\int_S p\mathbf{n} dS) dt, \tag{38}$$

$$(\int_V \rho \tilde{\mathbf{v}} dV)_{t=t_2} - (\int_V \rho \tilde{\mathbf{v}} dV)_{t=t_1} = 0 \tag{39}$$

and

$$(\int_V \rho \frac{\bar{\mathbf{v}} \cdot \bar{\mathbf{v}} + \tilde{\mathbf{v}} \cdot \tilde{\mathbf{v}}}{2} dV)_{t=t_2} - (\int_V \rho \frac{\bar{\mathbf{v}} \cdot \bar{\mathbf{v}} + \tilde{\mathbf{v}} \cdot \tilde{\mathbf{v}}}{2} dV)_{t=t_1} = -\int_{t_1}^{t_2} (\int_S \rho \bar{\mathbf{v}} \cdot \mathbf{n} dS) dt, \qquad (40)$$

$$(\int_V \rho \bar{\mathbf{v}} \cdot \tilde{\mathbf{v}} dV)_{t=t_2} - (\int_V \rho \bar{\mathbf{v}} \cdot \tilde{\mathbf{v}} dV)_{t=t_1} = 0 \qquad (41)$$

The law of conservation of mass can be written in the following form

$$(\int_V \rho dV)_{t=t_2} - (\int_V \rho dV)_{t=t_1} = 0 \qquad (42)$$

Here V is an arbitrary volume of fluid bounded by a surface S, **n** is a unit vector of the normal to S, $t_1, t_2$ are two arbitrary instants of time.

Since we are dealing with an incompressible inviscid fluid, the thermal energy is not included in the conservations laws. It should be noticed that pressure and density remain singlevalued and that is why they are referred to the whole particle rather than to each half-particle.

If all the variables exist and they are differentiable within the selected volume and time interval, the differential form of the governing equations (33)-(35) can be derived from the conservation laws (36)-(38). However, Eqs. (33)-(35) cannot be applied to describe formation and propagation of velocity jumps that can occur in an inviscid fluid, and that is the reason to turn to the conservation laws (36)-(38). We will be interested in behavior of tangential jumps of velocities since only that type of jumps leads to turbulence. In a singlevalued model represented by the Euler's equations tangential jumps do not propagate: they are unstable, and their postinstability behavior – turbulent mixing – cannot be described without additional experiment-based parameters such as mixing length. We will show here that in a doublevalued setting, the surface of a tangential jump splits into two separate surfaces that move away from each other **remaining stable** and propagating a turbulent mixing region.

**6. Conditions of dynamical compatibility.**

We start with the law of mass conservation and apply Eq. (42) to the surface S at its initial position prior to splitting in two surfaces moving in opposite directions

$$\rho(\lambda_1 + \lambda_2) = 0 \quad at \quad t = 0 \qquad (43)$$

Here $\lambda_1$ and $\lambda_2$ are the speeds of propagations of surfaces of discontinuities $S_1$ and $S_2$ originated from the split of the surface S. Since the momentum and energy conservation laws were applied to each half-particle separately, we can adopt classical derivation of the conditions of dynamical compatibility at the surface of the velocity jump following from Eqs. (36) and (37). We start with Eq. (36) and apply it to the surface $S_1$ moving upward with a speed $\lambda_1$

$$\rho \lambda_1 (\mathbf{v}_1^+ - \mathbf{v}_1^-) = -(p_1^+ - p_1^-) \mathbf{n}_1 \qquad (44)$$

$$\rho \lambda_1 (\mathbf{v}_2^+ - \mathbf{v}_2^-) = -(p_2^+ - p_2^-) \mathbf{n}_1 \qquad (45)$$

Here the superscripts denote the values of the parameters from both sides of the surface of jump, and subscripts denote the number of a half-particle.

In the same way, Eq. (37) leads to the following

$$\rho\lambda_1(\frac{\mathbf{v}_1^+\cdot\mathbf{v}_1^+}{2}-\frac{\mathbf{v}_1^-\cdot\mathbf{v}_1^-}{2})=(p_1^+\mathbf{v}_1^+-p_1^-\mathbf{v}_1^-)\cdot\mathbf{n}_1 \qquad (46)$$

$$\rho\lambda_1(\frac{\mathbf{v}_2^+\cdot\mathbf{v}_2^+}{2}-\frac{\mathbf{v}_2^-\cdot\mathbf{v}_2^-}{2})=(p_2^+\mathbf{v}_2^+-p_2^-\overline{\mathbf{v}}_2)\cdot\mathbf{n}_1 \qquad (47)$$

Eqs. (44)-(47) can be expressed via the mean/fluctuations velocities

$$\rho\lambda_1(\overline{\mathbf{v}}^+-\overline{\mathbf{v}}^-)=-(p^+-p^-)\mathbf{n}_1 \qquad (48)$$

$$\rho\lambda_1(\tilde{\mathbf{v}}^+-\tilde{\mathbf{v}}^-)=0 \qquad (49)$$

$$\rho\lambda_1(\frac{\overline{\mathbf{v}}^+\cdot\overline{\mathbf{v}}^++\tilde{\mathbf{v}}^+\cdot\tilde{\mathbf{v}}^+}{2}-\frac{\overline{\mathbf{v}}^-\cdot\overline{\mathbf{v}}^-+\tilde{\mathbf{v}}^-\cdot\tilde{\mathbf{v}}^-}{2})=(p^+\overline{\mathbf{v}}^+-p^-\overline{\mathbf{v}}^-)\cdot\mathbf{n}_1 \qquad (50)$$

$$\rho\lambda_1(\overline{\mathbf{v}}^+\cdot\tilde{\mathbf{v}}^--\overline{\mathbf{v}}^+\cdot\tilde{\mathbf{v}}^-)=0 \qquad (51)$$

It worth noticing that the first four conditions hold at the moving surfaces, respectively, while the last condition is referred to the original surface prior to its split in two.

**7. Mixing.**

*7.1. General remarks.*

Mixing is one of the specific phenomena generated by turbulence. The classical approach to this phenomenon requires the introduction of additional experimentally based variables such as Prandtle's mixing length. In order to clarify the "machinery" of mixing in a doublevalued EE model of incompressible inviscid fluid, we will start with a qualitative description of mixing on a local level that involves contact of two particle exchanging with their half-particles, Fig.2.

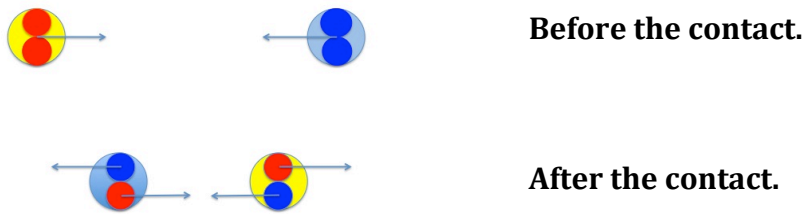

Before the contact.

After the contact.

 **Figure 2. Mixing as a transition to doublevaluedness.**

As shown in the Fig 2, before the contact, two particles move toward each other with the velocities ±v. Their half-particles move with the same velocities, so the velocity field is singlevalued. After the contact, the particles exchange their half's in such a way that each half-particle continues the motion with the same velocity, but within the "body" of another particle. Therefore the mixing propagates with the speed $\pm \mathbf{v}$ in both directions. However the velocity field becomes doublevalued: each particle has two half-particles moving in opposite directions with the same speed, and in our terminology, these are the fluctuation velocities. Obviously the mean velocity of each particle in the mixing zone is zero. The qualitative picture is similar in case of an oblique contact when the particles have different velocities before mixing: the mean velocity could be non-zero, and vertical fluctuation may occur in addition to the horizontal ones.

It should be emphasized that the doublevalued model permits reflection of the fluid from a rigid wall. Indeed, the boundary conditions can be formulated as following

$$\overline{\mathbf{v}} \cdot \mathbf{n} = 0, \quad (\tilde{\mathbf{v}}_1 + \tilde{\mathbf{v}}_2) \cdot \mathbf{n} = 0, \quad i.e. \quad \tilde{\mathbf{v}}_1 \cdot \mathbf{n} = -\tilde{\mathbf{v}}_2 \cdot \mathbf{n} \tag{52}$$

where $\mathbf{n}$ is the unit normal to the rigid wall. As follows from Fig. 2, the particle approaching the wall is "mixing" with a virtual particle symmetric with respect to the wall and moving to the same point of contact.

There are, at least, two different sources of mixing: instability of a tangential jump of velocity, and sharp angles streamlined at the boundaries. In this paper, we will concentrate only on the first source.

*7.2. Mixing triggered by a tangential velocity jump.*

Let us consider a surface of a tangential jump of velocity $(\mathbf{v}_2 - \mathbf{v}_1) \cdot \boldsymbol{\tau}$ in a horizontal unidirectional flow of an inviscid incompressible fluid assuming that this surface is not penetrated by the mean velocity $\overline{\mathbf{v}}$ of the double-valued velocity field i.e.

$$\overline{\mathbf{v}} \cdot \mathbf{n} = \overline{\mathbf{v}}_n = 0 \tag{53}$$

where $n$ and $\boldsymbol{\tau}$ are the normal and tangent to the surface of discontinuity, respectively,. Fig.3

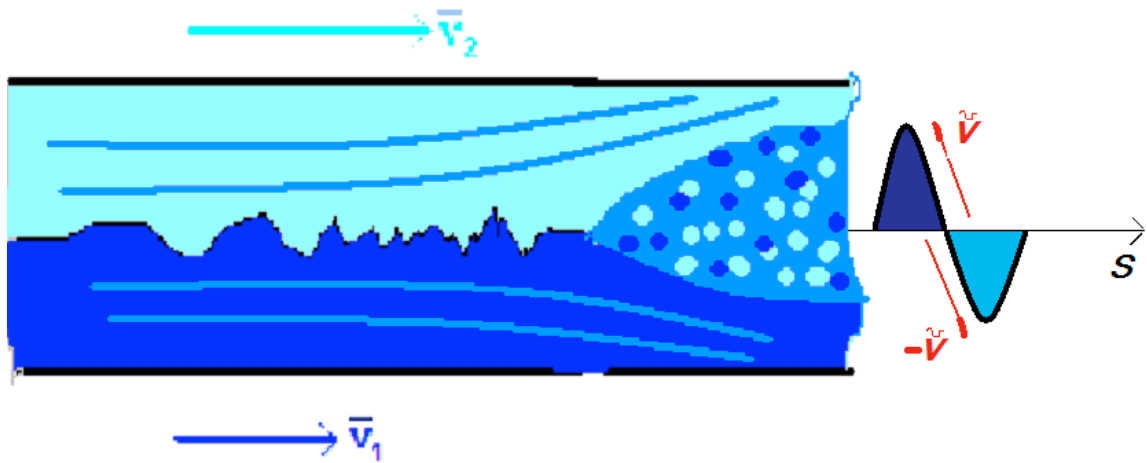

**Figure 3.** . **Mixing triggered by a tangential velocity jump.**

In classical (singlevatued) model, the surface $S$ is unstable, and the mixing followed this instability is described with help of ***additional parameters*** (such as mixing length) found from ***experiments***. As will be shown below, in the doublevalued setting, the surface $S$ splits in two half's that remain stable and move in opposite directions outlining the mixing zone.

We will start out analysis with application of dynamical compatibility conditions at the surface of discontinuity *S* (see Eqs. (44)-(51)).

First we turn to Eq. (43) that follows from the law of mass conservation at the surface of discontinuity. Since the speed of propagation of a tangential jump of velocity is equal to the normal component of the velocity of the fluid, i.e.

$$\lambda = \tilde{v}_n \tag{54}$$

Eq.(43) takes the form

$$\tilde{v}_n^+ + \tilde{v}_n^- = 0 \quad at \quad t = 0 \tag{55}$$

where $\tilde{v}_n^+, \tilde{v}_n^-$ are normal components of velocity fluctuations applied to different half's of the surface S. Thus the surface *S* splits in two half's that propagate in opposite directions with the same characteristic speeds $\pm \tilde{v}_n$ that do not depend upon the values of the parameters transferred. The same characteristic speeds were obtained in [3] for propagation of weak discontinuities of tangential component of velocities, or jumps of vortices. Such a coincidence takes place when strong discontinuities do not form shock waves.

It should be noticed that similar split of the surface of a *normal* velocity jump can occur in compressible *singlevalued* models when the conditions of dynamical compatibility do not hold; in this case the separated surfaces can move with different speeds, and the moment of the split is associated with an explosion. Although in our case the mechanism of the separation is different, but as follows from Eq. (55), the speed of propagation is characterized by a **normal** jump. The same jump characterizes the normal component of the fluctuations.

As noticed above, the conditions of dynamical compatibility (39)-(42) that were derived for a doublevalued model are valid for t>0 and
cannot be applied to the very first moment of contact between the two flows of fluid because at this moment the surface S is still singlevalued, and its behavior is governed by the Euler equation. It should be recalled that the doublevalued EE model must be applied **only** when the singlevalued model being applied to the same problem fails. If this failure results from instability (in the class of singlevalued functions), the information about the onset of this instability must be included into Eqs. (43)-(51). That is why we have to turn to the classical solution of the tangential jump of velocity that was discussed in [3]. As shown there, the solution of the dynamics of the surface *S* in Fig. 3 subject to initial conditions

$$U^*_0 = \frac{1}{\lambda_0} e^{-\lambda_0 S i}, \quad at \quad t = 0 \tag{56}$$

contains the fastest growing term

$$U = \frac{1}{\lambda_0} e^{\lambda_0 |Im \lambda'| \Delta t} \sin \lambda_0 S + ..., \quad \lambda_0 \to \infty \tag{57}$$

where *U* is the vertical displacement of *S*, and the characteristic roots are

$$\lambda' = \frac{1}{2}[(\mathbf{v}_2 + \mathbf{v}_1) \pm i(\mathbf{v}_2 - \mathbf{v}_1)] \cdot \mathbf{\tau} \tag{58}$$

Therefore

$$|\frac{\partial U}{\partial t}|_{max} = \frac{1}{2}(\mathbf{v}_2 - \mathbf{v}_1) \cdot \mathbf{\tau} \quad at \quad t=0 \tag{59}$$

and

$$\lambda_1 = -\lambda_2 = \frac{1}{2}(\mathbf{v}_2 - \mathbf{v}_1) \cdot \mathbf{\tau} \quad at \quad t=0 \tag{60}$$

Thus Eq. (60) defines the characteristic speed of separation of the surface of the tangential jump of velocity at $t=0$.

The next dynamical compatibility conditions must be applied to each of the moving surfaces separately. We will start with Eqs. (44) and (45) that express conservation of momentum. Projecting Eq.(44) on the normal $\mathbf{n}_1$ to the moving upward surface $S_1$, with reference to Eqs. (52) and (53), one obtains

$$\frac{1}{2}\rho[\tilde{v}_{n_1}^2]_1 = -[p]_1 \tag{61}$$

Similar condition holds for another moving downward surface

$$\frac{1}{2}\rho[\tilde{v}_{n_2}^2]_2 = -[p]_2 \tag{62}$$

Here the square brackets denotes a jump of the corresponding variable. As follows from Eqs. (61) and (62), the pressure has the negative change when it crosses over each moving half of the surface S outward the mixing zone, i.e. the pressure in the mixing zone is lower than outside of this zone.

Projections of Eq. (44) on the tangents to the propagating surfaces, relates the tangential jumps of the mean and fluctuation components of the velocity and the initial velocity jump on each moving surface

$$\{(\overline{\mathbf{v}} + \tilde{\mathbf{v}}) \cdot \mathbf{\tau}\}_1 = \mathbf{v}_2 \quad or \quad \overline{v}_{\tau_1} + \tilde{v}_{\tau_1} = v_2 \tag{63}$$

$$\{(\overline{\mathbf{v}} - \tilde{\mathbf{v}}) \cdot \mathbf{\tau}\}_2 = \mathbf{v}_1 \quad or \quad \overline{v}_{\tau_2} - \tilde{v}_{\tau_2} = v_1 \tag{64}$$

This equalities demonstrate that the initial tangential jump of velocity that leads to the instability of the surface $S$, is eliminated by occurrence of tangential components of velocity fluctuations, and that suppresses the instability of the same motion as soon as we move to the doublevalued model. The stabilization effect of the fluctuations was discussed in details in[3].

Since we consider inviscid incompressible fluid, the thermal energy is not included, and therefore, the law of conservation of energy holds as soon as the laws of mass and momentum conservation do.

The dynamical compatibility equations considered above have not defined the characteristic speed of propagation of the surfaces of discontinuities, and in order to complete the description of the propagation of an initial tangential jump of velocity, one has to invoke the differential equations (33)-(35). For that purpose, let us choose the system of Cartesian coordinates and direct X along the surface S, and Y normal to this surface. Before projecting Eqs. (33)-35) on these axes, we will make a simplifying, but obvious assumption that all the variables do not depend upon $x$. Then we arrive at a system of four equations with respect to four unknown $\bar{v}_{\tau_1}, \tilde{v}_{\tau_1}, \tilde{v}_{n_1},$ and $p_0$

$$\frac{\partial \bar{v}_{\tau_1}}{\partial t} + \tilde{v}_{n_1}\frac{\partial \tilde{v}_{\tau_1}}{\partial y} = 0 \tag{65}$$

$$\frac{\partial \tilde{v}_{\tau_1}}{\partial t} + \tilde{v}_{n_1}\frac{\partial \bar{v}_{\tau_1}}{\partial y} = 0 \tag{66}$$

$$\frac{\partial \tilde{v}_{n_1}}{\partial t} = 0 \tag{67}$$

$$\tilde{v}_{n_1}\frac{\partial \tilde{v}_{n_1}}{\partial y} = -\frac{1}{\rho}\frac{\partial p}{\partial y} \tag{68}$$

These equations are defined within the area bounded by the $X$ axis and the surface of discontinuity propagating upwards, i.e.

$$0 < y \le y_{S_1} \tag{70}$$

They are to be solved subject to the following initial/ boundary condition

$$\bar{v}_{\tau_1} + \tilde{v}_{\tau_1} = v_2 \quad at \quad 0 < y \le y_{S_1} \tag{71}$$

$$\bar{v}_{\tau_1} = \frac{1}{2}(v_1 + v_2) \quad at \quad t = 0, \quad y = 0 \tag{72}$$

$$\tilde{v}_{\tau_1} = \frac{1}{2}(v_2 - v_1) \quad at \quad t = 0, \quad y = 0 \tag{73}$$

$$\tilde{v}_{n_1} = \frac{1}{2}(v_2 - v_1) \quad at \quad t = 0, y = 0 \tag{74}$$

First of all, the pressure $p$ can be expressed via $\tilde{v}_{n_1}$ from Eq. (68)

$$p = p_0 - \rho\frac{\tilde{v}_{n_1}^2}{2} \tag{75}$$

and therefore, Eq. (68) can be excluded from further considerations.
As follows from Eq. (67), the characteristic speed (74) remains constant at the whole mixing zone, i.e.

$$\tilde{v}_{n_1} = \frac{1}{2}(v_2 - v_1) = \tilde{v}_0 = const. \quad at \quad 0 < y \le y_{S_1} \tag{76}$$

Then elimination of fluctuation velocity from Eqs. (65) and (66) leads to a trivial hyperbolic PDE with respect to the mean velocity

$$\frac{\partial^2 \bar{v}_{\tau_1}}{\partial t^2} - \tilde{v}_0^2 \frac{\partial^2 \bar{v}_{\tau_1}}{\partial t^2} = 0, \quad at \quad 0 < y \le y_{S_1} \tag{77}$$

By similar transformation, the same equation can be obtained for the fluctuation velocity

$$\frac{\partial^2 \tilde{v}_{\tau_1}}{\partial t^2} - \tilde{v}_0^2 \frac{\partial^2 \tilde{v}_{\tau_1}}{\partial t^2} = 0, \quad at \quad 0 < y \leq y_{S_1} \qquad (78)$$

Solutions of these equations have the form of travelling waves

$$\overline{v}_{\tau_1} = \overline{f}(y - \tilde{v}_0 t) \qquad (79)$$

$$\tilde{v}_{\tau_1} = \tilde{f}(y - \tilde{v}_0 t) \qquad (80)$$

Here we consider only the waves propagating upward, since the problem is symmetric with respect of $X$, and the lower part of the mixing zone (y<0) do not have to be treated separately.

The functions $\overline{f}$ and $\tilde{f}$ are found from the conditions (72) and (73), respectively

$$\overline{f}(y = \tilde{v}_0 t) = \frac{1}{2}(v_1 + v_2) \qquad (81)$$

$$\tilde{f}(y = \tilde{v}_0 t) = \frac{1}{2}(v_2 - v_1) \qquad (82)$$

Indeed, as follows from Eqs. (76) and (67)

$$\frac{\partial \overline{v}_{\tau_1}}{\partial y} = 0 \quad at \quad 0 < y \leq y_{S_1} \qquad (83)$$

and therefore, with reference to Eq. (71)

$$\frac{\partial \tilde{v}_{\tau_1}}{\partial y} = 0 \quad at \quad 0 < y \leq y_{S_1} \qquad (84)$$

Thus the solution (77), (78) describes propagation of the mixing zone upward with the constant characteristic speed (76) transporting constant mean and fluctuation components of the velocity field (79) and (80) respectively.

Similar solution can be obtained for the mixing zone propagating downward.

As follows from Eq. (75), the pressure in the whole mixing zone is constant

$$p = p_0 - \rho \frac{\tilde{v}_0^2}{2} = p_0 - \frac{1}{4}\rho(v_2 - v_1) \qquad (85)$$

but it is lower than in the unperturbed zone.

*7.3. Reflection from the boundaries.*

So far we have considered an unbounded propagation of the mixing zone. Let us now assume that the flow is bounded by smooth solid surfaces from above and below. We recall that the doublevalued model under consideration allows fluctuations to be reflected from solid smooth surface along the normal to this surface without loss of energy. This follows from the property that the divergence of fluctuation velocity is not bounded

$$\nabla \cdot \tilde{\mathbf{v}} \neq 0 \qquad (86)$$

(see the comments to Eq. (52) and Fig. 2)).

An analytical description of reflection of vertical fluctuations from the boundaries is similar to reflection of waves in a vibrating string from the fixed ends since the governing equations for both phenomena are identical,

(see Eqs. (77) and (78)). The reflected fluctuations have the same value, but they do not transport any discontinuities any more.

*7.4. Interpretation of the solution.*

The solution looks very transparent (see Figure 4): the mixing zone propagates upward and downward with the constant speed equal to the half of the original velocity jump (see Eq. (76)); the mean velocity in the mixing zone is equal to that at the surface of contact (see Eq. (72)). The mean velocity field is accompanied by horizontal and vertical velocity fluctuations in the mixing zone: they are constant and equal to the half of the original velocity jump (see Eqs. (73) and (76)); in addition to that, the vertical fluctuations are reflected from the boundaries without loss of energy (see Eq. (52)).

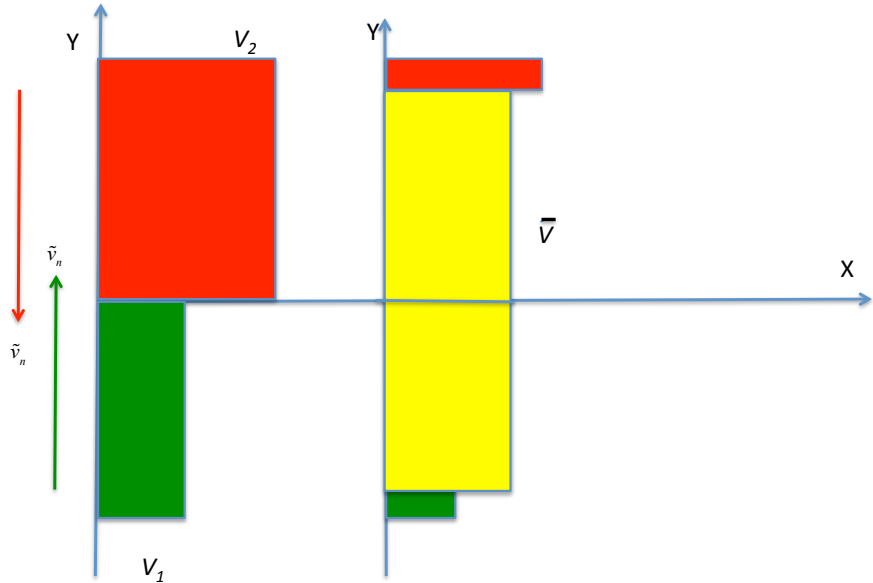

**Figure 4. Propagation of the mixing zone.**

Finally, the pressure in the mixing zone is reduced by the specific kinetic energy of vertical fluctuation (see Eq. (85)).

Let us compare this solution with the classical solution given by Prandtl, [7,8]. Keeping our notations, the Prandtl solution is presented as

$$\bar{v}_{\tau_1} = \frac{1}{2}(v_1 + v_2) + \frac{1}{2}(v_2 - v_1)[\frac{3}{2}(\frac{y}{b}) - \frac{1}{2}(\frac{y}{b})^3] \qquad (87)$$

$$b = \frac{3}{2}\beta^2(v_2 - v_1)t \tag{88}$$

A formal comparison demonstrates that for large times the solution (87) for the mean velocity in the mixing zone becomes identical to the solution (79) if the experimentally found coefficient is the following

$$\beta^2 = \frac{2}{3} \tag{89}$$

The rest of the parameters characterizing the mixing zone cannot be compared since Prandtl did not include velocity fluctuations and pressure in his model.

Let us turn to a qualitative comparison of these solutions.

The Prandtl model is based upon the Reynolds equations in which the physical viscosity is ignored. Nevertheless the ***parabolic*** type of these equations is artificially preserved by an experimentally based closure that includes the mixing length. That is why Prandtl could not consider the speed of propagation of the mixing zone as a characteristic one: he ***experimentally*** proved that this speed is constant. Despite several inconsistencies in the Prandtl model, it should be recognized as the first model of mixing that has many engineering applications.

The governing equations of the doublevalued model proposed above are of a ***hyperbolic type***, and therefore, it allows one to treat velocity jumps as strong discontinuities propagating with characteristic speeds. That creates a closed system of equations that does not require any experimentally based additions. Despite simplicity of the solution, it defines horizontal and vertical components of velocity fluctuations as well as the pressure in the mixing zone that the Prandtl solution did not define. However we have to emphasize that the solution proposed above is valid only for zero viscosity: strictly speaking, it can be applied only to superfluids (liquid helium, and some of Bose-Einstein condensates). Indeed in fluids with non-zero viscosity, no matter how small it is, a finite tangential jump of velocity would cause an unbounded shear stress. That is why for classical fluids no-slip condition at surfaces of tangential jumps of velocities (including rigid boundaries) must be enforced, and the doublevalued model should be applied only beyond the corresponding boundary layer (which width is inversely proportional to the Raynolds number), while a connection between the laminar motion within this layer and the turbulent motion beyond it is to be implemented by utilizing instability of the boundary layer for the initial/boundary conditions of the turbulent flow (see Eqs. (56)-(60)). Therefore the comparison with experiments performed on a real fluid (that always has some viscosity) may show a discrepancy in the areas around tangential jumps. Nevertheless the values of idealized models are demonstrated by the discovery of sound and shock waves: these fundamental phenomena can exist only in ideal (inviscid) model of fluid since, strictly speaking, viscous models cannot have discontinuities of the velocity field. And this is another angle to view the difference between the proposed and the Prandtl's solutions.

*7.5. Comments to logarithmic laws.*

An explanation of a sharp difference between laminar and turbulent velocity profiles of shear flows about an unbounded wall was always a test of a theory of turbulence. So far the derivation of the logarithmic profile of turbulent motions has been based upon experimentally found additional parameters associated with the mixing length, [7,8]. We will propose here a qualitative explanation of this law in context of the propagation of a mixing zone in the Prandtl's problem discussed above. Let us select horizontal and vertical Cartesian axes *X* and *Y*, respectively and consider a plane horizontal laminar flow about an unbounded wall, ignoring volume forces. In this case, any two vertical cross-sections will be identical, and all the derivatives with respect to *x* will be zero. The laminar profile of velocity is given by the straight line, [10]

$$v_x = \frac{\sigma_0 \rho}{\nu} y, \quad p = p_0 = const. \tag{90}$$

where $v_x, \nu, \sigma_0$ are horizontal velocity, kinematical viscosity, and shear stress at the wall, respectively.

Assume now that the horizontal velocity increases such that the Reynolds number becomes supercritical, and therefore, the velocity profile (90) becomes unstable, i.e. any small disturbance of the velocities grows exponentially. One of the most visible results of this instability is mixing. First we have to evaluate the thickness of the boundary layer that is still laminar. Since for our derivation, the exact value of thickness of this layer is insignificant, we will use an approximation, [10],

$$\delta = \alpha \nu \sqrt{\frac{\rho}{\sigma_0}} \tag{91}$$

Similar approximation can be applied for the velocity on the boundary between the laminar and turbulent flows

$$v_x^0 = \alpha \sqrt{\frac{\sigma_0}{\rho}} \tag{92}$$

In these equations, $\alpha$ is a dimensionless coefficient that is insignificant for our discussion as well.

Next consider the area of the flow above the boundary layer

$$y \geq \delta \tag{93}$$

and assume that there is a tangential jump of velocity at the boundary between the laminar and turbulent flows as a result of instability. Then the mixing process will start that qualitatively is the same as that described in the previous sub-section, with the only difference that now the velocity **v₂** depends linearly upon *y*

$$v_2 = v_x^0 + \Delta v + \frac{\sigma_0 \rho}{\nu} y \tag{94}$$

where $\Delta v$ is the velocity jump.

The solution of the transition to turbulence can be formally described by Eqs. (81) and (82)

$$\bar{f}(y = \tilde{v}_0 t) = \frac{1}{2}[v_1 + v_2(y)] \tag{95}$$

$$\tilde{f}(y = \tilde{v}_0 t) = \frac{1}{2}[v_2(y) - v_1] \tag{96}$$

where $v_2$ is given by Eq. (94).

Let us now turn to the measurement problem. Most of the velocity sensors actually measure velocity via the pressure. But in presence of fluctuations, the pressure includes not only the contribution of the mean velocity, but the fluctuations as well. That means that such sensors measure some "effective" mean velocity

$$\bar{\bar{v}} = \sqrt{\bar{v}^2 + \tilde{v}^2} \tag{97}$$

Substituting in Eq. (97) the expressions from Eqs. (95) and (96) with reference to Eq. (94) on arrive at a profile $\bar{\bar{v}}(y)$ that qualitatively similar to a logarithmic law, Fig. 5. It should be noticed that in many particular cases the logarithmic law is replaced by more accurate power laws like Darcy law, [10].

Thus it can be suggested that turbulence does not change the mean velocity profile: it rather stabilizes it by horizontal fluctuations. It is the "effective "mean, or measured velocity profile that deviate from the laminar profile.

It should be emphasized that in the classical models, the mean and the "effective" mean velocities are the same.

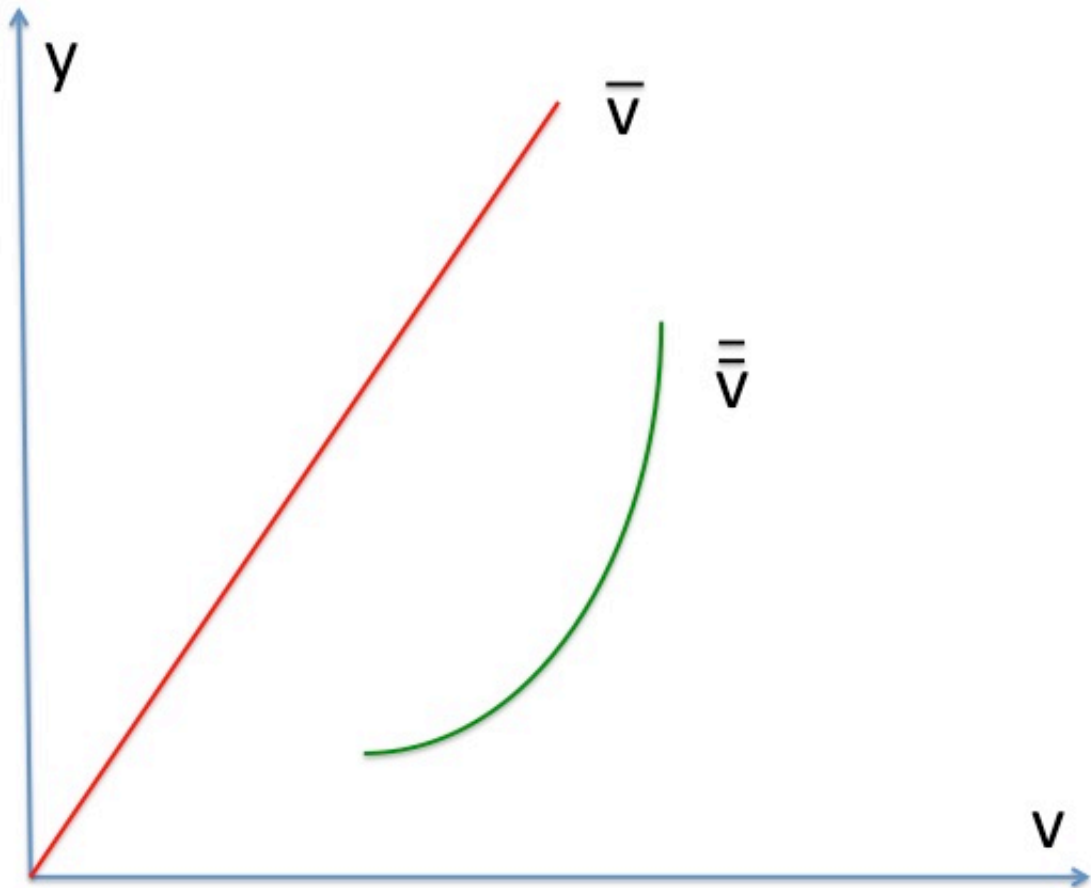

**Figure 5. Turbulent profiles of mean and "effective" mean velocities.**

**8**. **Modified Navier-Stokes equations.**

Starting with this section, we move from Euler to Navier-Stokes equations. The rationale of that is the following: A developed turbulence is characterized by mean and fluctuation velocities, while fluctuations can be divided in two classes: small and large scale fluctuations. The small-scale fluctuations ( $Re \approx 1$ ) are responsible for dissipation of mechanical energy, and practically they do not affect the general picture of motion since their amplitudes are small compare to mean velocities. The large-scale fluctuations ( $Re \to \infty$ ) are sizable with the mean velocities, and they significantly contribute to the motions. These properties suggest that the general picture of turbulence is better captured by the Euler rather than Navier-Stokes equation, and in particular, by EE equations, regardless of whether the underlying pre-instability laminar flow is viscous or non-viscose. The EE equations (1-3) actually implement the approach described above. However the EE

model has the following limitation: it is exact only for zero viscosity

$$\nu = 0 \tag{98}$$

while for even an infinitesimal viscosity

$$0 < \nu \to 0 \tag{99}$$

the EE model can be considered only as an approximation.

The explanation of this discontinuous dependence of the solution on the viscosity when it passes from positive value to zero is due to the change of the PDE from a parabolic to a hyperbolic type at the point (98) since at this point the highest order derivatives of the velocity vanish. But since most of the real fluids (except of superfluids) belong to the type (99), a viscous modification of EE equations seems useful. This means that a fluid flow is subdivided on three area, [10]: the first area is a laminar sub-layer that is characterized only by the physical viscosity; the second one is a transitional area characterized by both physical and turbulent viscosity; and finally, the third area is turbulent core characterized only by the turbulent viscosity. In our setting, the first area is described by the classical Navier-Stokes equations, the third area – by EE equations, and the second area is supposed to be modeled by modified Navier-Stokes equations that are characterized by a doublevalued velocity field, but preserve the contribution of the physical viscosity without violation of boundedness of stresses. The modification that combines the Navier-Stokes and EE equations is the subject of the next sections. We will consider here both incompressible and compressible fluids.

**9. Semi-viscose incompressible fluid.**

*a. Model derivation*

With reference to [1], we introduce a doublevalued velocity field

$$v(x_1) \neq v(x_2) \quad if \quad x_1 = x_2 \tag{100}$$

Such an idealization means that two different particles with different velocities can appear at the same point of space without causing any physical inconsistency, and that is possible only due to the absence of shear stresses, (this property will be verified later). Actually we arrive at two superimposed continua, and each of them can be described by slightly modified Navier-Stokes equations (see Fig.1). In case of incompressible fluid the governing equations are

$$\frac{\partial \mathbf{v}_1}{\partial t} + \mathbf{v}_1 \nabla \mathbf{v}_1 = \nu \nabla \cdot dev \nabla (\mathbf{v}_1 - \mathbf{v}_2) - \frac{1}{\rho} \nabla p + \mathbf{F} \tag{101}$$

$$\frac{\partial \mathbf{v}_2}{\partial t} + \mathbf{v}_2 \nabla \mathbf{v}_2 = \nu \nabla \cdot dev \nabla (\mathbf{v}_2 - \mathbf{v}_1) - \frac{1}{\rho} \nabla p + \mathbf{F} \tag{102}$$

$$\nabla \bullet (\mathbf{v}_1 + \mathbf{v}_2) = 0 \tag{103}$$

where ***F*** is external force per unit mass, and $\nu$ is kinematical viscosity.

It should be emphasized that the pressure *p*, the density $\rho$ as well as the divergence of velocity must remain single-valued as in the EE model. The modified Navier-Stokes (MNS) model differs from the EE model by additional viscous terms that represent **internal** friction between two half-particles. As will be shown below, this friction does not affect the center of mass of the entire particle and does not generate a shear stress. Indeed, adding up Eqs.(101) and (102), and subtracting them from one another one obtains respectively

$$\frac{\partial \overline{\mathbf{v}}}{\partial t} + \overline{\mathbf{v}}\nabla\overline{\mathbf{v}} + \tilde{\mathbf{v}}\nabla\tilde{\mathbf{v}} = -\frac{1}{\rho}\nabla p + \mathbf{F} \tag{104}$$

$$\frac{\partial \tilde{\mathbf{v}}}{\partial t} + \overline{\mathbf{v}}\nabla\tilde{\mathbf{v}} + \tilde{\mathbf{v}}\nabla\overline{\mathbf{v}} = \nu\nabla \cdot dev\nabla\tilde{\mathbf{v}} \tag{105}$$

Eq. (103) in new variables can be rewritten in the form

$$\nabla \bullet (\overline{\mathbf{v}}) = 0 \tag{106}$$

Thus we arrive at a closed system of two vector and one scalar equations with respect to two vectors, $\overline{\mathbf{v}}, \tilde{\mathbf{v}}$ and one scalar *p* unknowns. Eq. (104) that represents dynamics of the mean velocity is exactly the same as Eq. (33) of the EE model: it does not include viscose terms, and therefore, it is free of shear stresses. That protects the model from unbounded stresses at tangential jumps of the mean velocity, thereby justifying the introduction of the doublevalued velocity field. However since the mean and fluctuation velocities are coupled, the latter restrict propagation of discontinuities of the mean velocities. We will consider this effect in more details below

*b. General properties of the model.*

By general properties of the model we understand such its properties that do not depend upon initial/ boundary conditions explicitly. The most important property of this kind is a type (hyperbolic/parabolic) of the PDE under consideration. Since the type of PDE is closely associated with existence and propagation of discontinuities of the model's variables, we will start with analysis the velocity field around a tangential jump of a mean velocity vortex. For that purpose project Eqs. (104) and (105) onto the Cartesian coordinates *X* , and *Y,* assuming that *Y* coincides with the normal to the surface of discontinuity. We will write these projections in terms of jumps of the first derivatives when the corresponding variable crosses the surface of discontinuity. Obviously in this case only the derivatives with respect to *y* will survive since the jumps of the derivatives with respect to *x* and *z* must be zero as follows from the conditions of kinematical compatibility at the surface of discontinuity,

$$[\partial/\partial n] = 0 \quad if \quad [\partial/\partial \tau] \neq 0 \tag{107}$$

and vice versa.

Then we arrive at a system of four equations with respect to four unknown $\bar{v}_\tau, \tilde{v}_\tau, \tilde{v}_n,$ and $p_0$

$$\frac{\partial \bar{v}_\tau}{\partial t} + \tilde{v}_n \frac{\partial \tilde{v}_\tau}{\partial y} = 0 \tag{108}$$

$$\frac{\partial \tilde{v}_\tau}{\partial t} + \tilde{v}_n \frac{\partial \bar{v}_\tau}{\partial y} = \nu \frac{\partial^2 \tilde{v}_\tau}{\partial y^2} \tag{109}$$

$$\frac{\partial \tilde{v}_n}{\partial t} = 0 \tag{110}$$

$$\tilde{v}_n \frac{\partial \tilde{v}_n}{\partial y} = -\frac{1}{\rho} \frac{\partial p}{\partial y} \tag{111}$$

As follows from Eq. (109),

$$[\frac{\partial \tilde{v}_\tau}{\partial t}] = -\lambda [\frac{\partial \bar{v}_\tau}{\partial y}] = 0 \tag{112}$$

where λ is the characteristic speed of propagation of a discontinuity; the condition (112) protects the right side of Eq.(109) from being unbounded. Turning now to Eq.(108), one observes that

$$[\frac{\partial \bar{v}_\tau}{\partial t}] = -\lambda [\frac{\partial \bar{v}_\tau}{\partial y}] = 0 \tag{113}$$

Hence

$$\lambda = 0 \quad if \quad [\frac{\partial \bar{v}_\tau}{\partial y}] \neq 0 \tag{114}$$

Thus the incompressible version of the MNS model is of a hyperbolic type when the fluctuation velocity are zero, and of a parabolic type otherwise.

*c. Transition to turbulence.*

The main purpose of the modification of the Navier-Stokes equations is to describe the transition to turbulence as a stable dynamical process by introducing the doublevalued velocity field while preserving some viscosity effects. In order to demonstrate a stable transition provided by MNS, let us start with a tangential velocity jump. Such jump is possible in MNS if velocity fluctuation are absent since than the model is degenerated into Euler's equations.

Let us consider a surface of a tangential jump of velocity $(\mathbf{v}_2 - \mathbf{v}_1) \cdot \boldsymbol{\tau}$ in a horizontal unidirectional flow of an inviscid incompressible fluid assuming that this surface is not penetrated by the mean velocity $\bar{\mathbf{v}}$ of the double-valued velocity field i.e.

$$\bar{\mathbf{v}} \cdot \mathbf{n} = \bar{v}_n = 0 \tag{115}$$

where $\mathbf{n}$ and $\boldsymbol{\tau}$ are the normal and tangent to the surface of discontinuity, respectively,. Fig.3.

In classical (singlevatued) model, the surface $S$ is unstable, and the mixing followed this instability is described with help of additional parameters (such as mixing length) found from experiments. As will be shown below, in the doublevalued setting, the surface $S$ splits in two half's that remain stable and move in opposite directions outlining the mixing zone. As in the inviscid model considered above, the propagation speed of the mixing zone is finite as it should be in hyperbolic PDE. In order to describe the transition to turbulence triggered by the tangential jump of velocity, we will apply Eqs. (108-111). However these equations cannot be applied to the very first contact between the two flows of fluid because at this moment the surface S is still singlevalued, and its behavior is governed by the Euler equation. It should be recalled that the doublevalued MNS model must be applied *only* when the singlevalued model being applied to the same problem fails. If this failure results from instability (in the class of singlevalued functions), the information about the onset of this instability must be included into Eqs. (100)-(103). That is why we have to turn again to the classical solution of the tangential jump of velocity that was discussed in [3] and turn to the solutions (56)-(59). That leads us to Eq.(60)

$$\lambda_1 = -\lambda_2 = \frac{1}{2}(\mathbf{v}_2 - \mathbf{v}_1) \cdot \boldsymbol{\tau} \quad at \quad t = 0 \tag{116}$$

Thus Eq. (116) defines the speed of separation of the surface of the tangential jump of velocity at $t=0$, as well as the initial value of the horizontal fluctuations. However unlike the inviscid doublevalued model described in the previous sections, this speed is not a characteristic for MNS equatins (108)-(111): it will play only the role of the boundary condition.

Let us turn to Eq. (110) that is decoupled from Eqs. (108) and (111). Its solution subject to the following initial and boundary condition

$$\tilde{v}_n(0,t) = \tilde{v}_n^0 = \frac{1}{2}(\bar{\mathbf{v}}_2 - \bar{\mathbf{v}}_1) \cdot \boldsymbol{\tau} = const \tag{117}$$

$$\tilde{v}_n(y,0) = 0 \tag{118}$$

is

$$\tilde{v}_n = \tilde{v}_n^0, \tag{119}$$

Thus the instability of the tangential jump of the velocity triggers the propagation of the mixing zone along the normal to this surface with constant speed exactly as in the EE model discussed in the section 7. The propagation is carried out by the velocity fluctuations normal to the surface of the jump. As follows from the solution (111), the motion is stable in the doublevalued class of functions.

Turning to Eq. (111), one finds the pressure distribution in the turbulent mixing zone that is the same as in the EE model discussed above.

$$p = p_0 - \frac{(\tilde{v}_n^0)^2}{2} \tag{120}$$

where $p_0$ is the initial pressure.

Since Eq. (110) is coupled with Eqs. (108) and (109) in a master-slave fashion, the vertical fluctuations trigger horizontal fluctuations as well as a change in the distribution of the mean velocities. The system reduced to a parabolic PDE of the third order and can be solved by separation of variables. The derivation of the solution will be given below.

*d. Velocity field in the mixing zone.*

The system of Eqs. (108) and (111) with reference to Eq.(119) reduces to the following equation

$$\frac{\partial^2 \tilde{v}_\tau}{\partial t^2} - (\tilde{v}_n^0)^2 \frac{\partial^2 \tilde{v}_\tau}{\partial y^2} = \nu \frac{\partial^3 \tilde{v}_\tau}{\partial t \partial y^2} \tag{121}$$

We will start with derivation of the general solution prior to formulating initial/boundary conditions. Applying the separation of variable technique, introduce new variables

$$\tilde{v}_\tau(t, y) = T(t)Y(y) \tag{122}$$

Substituting Eq. (121) into Eq. (122) one obtains two ODE

$$Y'' + \lambda Y = 0 \tag{123}$$

$$\ddot{T} + \lambda \nu \dot{T} + \lambda (\tilde{v}_n^0)^2 T = 0 \tag{124}$$

where $\lambda$ is a constant to be found from the boundary conditions.

First let us turn to Eq. (123) and write down its general solution

$$Y = C_1 e^{\sqrt{-\lambda} y} + C_2 e^{-\sqrt{-\lambda} y} \tag{125}$$

assuming that $\lambda$ is real and negative, $[\lambda] = \frac{1}{m^2}$.

Prior to formulating boundary and initial conditions, we have to specify the problem to be solved. For that purpose, consider a plane horizontal laminar flow about an unbounded wall, ignoring volume forces. In this case, any two vertical cross-sections will be identical, and all the derivatives with respect to *x* will be zero. The laminar profile of velocity is given by the straight line, [10]

$$v_x = b\hat{y}, \quad b = \frac{\sigma_0 \rho}{\nu}, \quad [b] = \frac{1}{\sec}, \quad p = p_0 = const. \quad \hat{y} = y + \delta, \tag{126}$$

where $v_x, \nu, \sigma_0$ and $\delta$ are horizontal velocity, kinematical viscosity, shear stress at the wall, and the thickness of the boundary layer, respectively.

Next consider the area of the flow above the boundary layer (see section 7.5)

$$y \geq 0 \tag{127}$$

and assume that there is a tangential jump of velocity at the boundary between the laminar and turbulent flows as a result of instability. Then the mixing process will start that qualitatively is the same as that described in the section 7.

Now we are ready to formulate the boundary condition for Eq. (125)

$$Y(0) = C_2 > 0 \tag{128}$$

$$Y(\infty) = 0 \tag{129}$$

The value of $C_2$ in Eq. (128) will be defined later. Based upon the boundary condition (129), one has to choose in Eq. (125) $C_1=0$, and therefore

$$Y = C_2 e^{-\sqrt{-\lambda} y} \tag{130}$$

that satisfies both boundary conditions.

Let us now turn to Eq. (124). Its general solution is

$$T = D_1 e^{A_1 t} + D_2 e^{A_2 t} \tag{131}$$

where

$$A_{1,2} = -\frac{1}{2}\lambda \nu \pm [\frac{1}{4}\lambda^2 \nu^2 - \lambda(\tilde{v}_n^0)^2]^{1/2} \tag{132}$$

In order to eliminate the unbounded component in Eq. (131), one has to choose

$$A = -\frac{1}{2}\lambda \nu - [\frac{1}{4}\lambda^2 \nu^2 - \lambda(\tilde{v}_n^0)^2]^{1/2} \tag{133}$$

since $A < 0$ at $\lambda < 0$.

Therefore Eq. (131) reduces to the following

$$T = D_1 e^{At} \tag{134}$$

Hence the solution of Eq. (121) can be written in the following form

$$\tilde{v}_\tau = C e^{-\sqrt{-\lambda} y + At}, \quad C = C_2 D_1 \tag{135}$$

where $C$ and $\lambda$ are to be found from the boundary and initial conditions.

From the following boundary/initial condition (see Eq. (116)

$$\tilde{v}_\tau(0,0) = \frac{1}{2} \Delta v_0 \tag{136}$$

where $\Delta v_0$ is the tangential jump of the velocity at $y=0$, $t > 0$ one obtains

$$C = \frac{1}{2} \Delta v_0 \tag{137}$$

In order to find $\lambda$, let us turn to Eq. (109) and write it for the velocity $\tilde{v}_\tau(0,0)$. Then with reference to Eq. (124) we can formulate another initial/boundary condition

$$\left(\frac{\partial \tilde{v}_\tau}{\partial t} - \nu \frac{\partial^2 \tilde{v}_\tau}{\partial y^2}\right)_{t=0, y=0} = -\Delta v_0 b \tag{138}$$

Now as follows from Eqs. (135) and (138)

$$A + \nu \lambda = -2b \tag{139}$$

whence

$$\lambda = -\frac{4b^2}{2\nu b + (\Delta v_0)^2}, \quad b = \frac{\sigma_0 \rho}{\nu} \tag{140}$$

Thus the solution of Eq. (121) that satisfies the boundary/initial conditions (136) and (138) is

$$\tilde{v}_\tau = \frac{1}{2}\Delta v_0 e^{-\sqrt{-\lambda}y + At}, \qquad (141)$$

where A and λ are given by Eqs. (133) and (140) respectively. As follows from this solution, the horizontal velocity fluctuations triggered by the tangential velocity jump $\Delta v_0$ are vanishing at infinity.

Let us turn to Eq. (108) and find the mean velocity profile

$$\overline{v}_\tau = B(y) + \frac{1}{2}\Delta v_0 \sqrt{-\lambda}\, e^{-\sqrt{-\lambda}y + At} \qquad (142)$$

where $B(y)$ is an arbitrary function to be found from the boundary condition

$$\overline{v}_\tau(0, y) = by \qquad (143)$$

i.e.

$$B(y) = by - \frac{1}{2}\Delta v_0 \sqrt{-\lambda}\, e^{-\sqrt{-\lambda}y} \qquad (144)$$

That leads to the mean velocity profile

$$\overline{v}_\tau = \frac{\sigma_0 \rho}{\nu} y - \frac{1}{2}\Delta v_0 \sqrt{-\lambda}\, e^{-\sqrt{-\lambda}y} + \frac{1}{2}\Delta v_0 \sqrt{-\lambda}\, e^{-\sqrt{-\lambda}y + At} \qquad (145)$$

As follows from Eq. (145), this profile gradually deviates from the straight-line laminar profile to the turbulent profile

$$\overline{v}_\tau = \frac{\sigma_0 \rho}{\nu} y - \frac{1}{2}\Delta v_0 \sqrt{-\lambda}\, e^{-\sqrt{-\lambda}y} \quad at \quad t \to \infty \qquad (146)$$

Combining Eqs. (141), (145) and (97), one arrives at the effective velocity $\overline{\overline{v}}(y)$ that qualitatively similar to the logarithmic law.

## 10. Semi-viscose compressible fluid

*a. Model derivation.*

For the compressible doublevalued velocity field model, the MNS equations governing turbulent motions can be written as an extension of Eqs. (93),(94) with the change of the continuity equation (95)

$$\frac{\partial \overline{\mathbf{v}}}{\partial t} + \overline{\mathbf{v}}\nabla\overline{\mathbf{v}} + \tilde{\mathbf{v}}\nabla\tilde{\mathbf{v}} = \frac{1}{\rho}(-\nabla p + \mathbf{F}) \tag{147}$$

$$\frac{\partial \tilde{\mathbf{v}}}{\partial t} + \overline{\mathbf{v}}\nabla\tilde{\mathbf{v}} + \tilde{\mathbf{v}}\nabla\overline{\mathbf{v}} = \nu\nabla \cdot dev\nabla\tilde{\mathbf{v}} \tag{148}$$

$$\frac{\partial \rho}{\partial t} + \nabla \bullet (\rho\mathbf{v}) = 0, \quad p = f(\rho, T) \tag{149}$$

Here $T$ is the temperature that is an additional unknown variable. For the closure we will need the heat balance equation

$$\rho\frac{di}{dt} = \frac{dp}{dt} + 2\rho\nu\dot{S}^2 \tag{145}$$

Here $i$ is the enthalpy

$$i = Jc_p T \tag{151}$$

where $J$ is the mechanical equivalent of heat, and $c_p$ is the specific heat, $\dot{S}$ is the deviator of the stress tensor

$$S_{ij} = \frac{\partial \tilde{v}_i}{\partial x_j} + \frac{\partial \tilde{v}_j}{\partial x_i} \quad at \; i \neq j, \quad S_{ii} = 0 \tag{152}$$

*b. General properties.*
The MNS model of compressible fluid, in addition to the properties of the incompressible model considered in the previous section, is characterized by a system of hyperbolic waves that transport normal jumps of parameters with the possibility of shock waves as in the Euler's equations. The possibility of these waves exists since the viscous components of stresses are defined by the **deviator** of the tensor of velocity gradient (see Eqs. (152)), and therefore, the normal components of velocities are not affected by the viscosity directly, (see Eq.(110)). It should be noticed that an important property of the proposed model is the coupling of the governing equations of motion with the thermal balance. Indeed as suggested in [10], introduction of an additional intermediate zone between a laminar boundary layer and inviscid turbulence is especially important for analysis of heat transfer. Obviously the doublevaluedness of the velocity field affects not only heat transfer processes, but also such fundamental invariants as speed of sound,[3], as well as shock waves characteristics. However, an analysis of thermal processes and shock waves propagation in turbulent flows is out of scope of this paper.

**11. Discussion and conclusion.**

The objective of this work is to prove that Newtonian mechanics is fully equipped for description of turbulent motions without help of experimentally obtained closures. Turbulence is one of the most fundamental problems in theoretical physics that is still unsolved. Although applicability of the Navier-Stokes equations as a model for fluid mechanics is not in question, the instability of their solutions for flows with supercritical Reynolds numbers raises a more general question: is Newtonian mechanics complete?

The problem of turbulence (stressed later by the discovery of chaos) demonstrated that the Newton's world is far more complex than those represented by classical models. It appears that the Lagrangian or Hamiltonian formulations do not suggest any tools for treating postinstability motions, and this is a major flaw of the classical approach to Newtonian mechanics. The explanation of that limitation was proposed in [9]: the classical formalism based upon the Newton's laws exploits additional mathematical restrictions (such as space–time differentiability, and the Lipschitz conditions) that are not required by the Newton's laws. The only purpose for these restrictions is to apply a powerful technique of classical mathematical analysis. However, in many cases such restrictions are incompatible with physical reality, and the most obvious case of such incompatibility is the Euler's model of inviscid fluid in which absence of shear stresses are not compensated by a release of additional degrees of freedom as required by the principles of mechanics. This paper is concentrated on elimination of this incompatibility by introduction of non-differentiable (multivalued) velocity field. It is started with a detailed derivation of the EE equations from the principle of virtual work, followed by introduction of integral form of the governing equations, and analysis of propagation of velocity jumps. The theory is applied to turbulent mixing and illustrated by propagation of mixing zone triggered by a tangential jump of velocity. A comparison of the proposed solution with the Prandtl's solution is performed and discussed. In the second part of this paper, the properties of extended Euler equations (EE equations) characterized by a doublevalued velocity field started in [3], are generalized to the Navier-Stokes equations for the regions of turbulent motions. The modified Navier-Stokes (MNS) equations differ from the EE equations by additional viscous terms in the governing equations of fluctuations, but not in the equations for the mean velocities, and that protects the MNS model from unbounded stresses since the shear stresses are still zero. The model does not require any closures since the number of equations is equal to the number of unknowns. Special attention is paid to transition from laminar to turbulent state. The analytical solution for this transition demonstrates the turbulent mean velocity profile that qualitatively similar to celebrated logarithmic law.

**References.**